\documentstyle[sprocl]{article}
\input{psfig}
\def\Journal#1#2#3#4{{#1} {\bf #2}, #3 (#4)}

\def\be{\begin{equation}}
\def\ee{\end{equation}}
\def\bea{\begin{eqnarray}}
\def\kms{km~s$^{-1}$}
\def\cm2{cm$^{-2}$}
\def\eea{\end{eqnarray}}
\def\qn{QSO\,1937$-$1009\,}
\def\q2{QSO\,1009+2956\,}
\def\Lya{Lyman-$\alpha$~}
\def\Lyb{Lyman-$\beta$~}
\def\Lyg{Lyman-$\gamma$~}
\def\Lyd{Lyman-$\delta$~}

\newcommand {\apgt} {\ {\raise-.5ex\hbox{$\buildrel>\over\sim$}}\, } \newcommand
{\aplt} {\ {\raise-.5ex\hbox{$\buildrel<\over\sim$}}\, }

\begin{document}

\title{THE COSMOLOGICAL BARYON DENSITY
FROM DEUTERIUM IN QSO SPECTRA}

\author{D. TYTLER and S. BURLES}

\address{CASS 0111, University of California San Diego, La Jolla, CA 92093-0111,
USA. tytler@ucsd.edu, scott@cass154.ucsd.edu}

\maketitle\abstracts{
The primordial D/H ratio now provides the best measure of the 
cosmological density of baryons.
We describe in detail how we deduce the D/H ratio from absorption
lines in the spectra of quasars, and we present our first
two measurements of D/H in different QSOs, which agree to within
the random errors. We describe how we correct D/H for
blending with Lyman alpha forest H lines. For our two QSOs these corrections 
are small, because the D lines are narrow, deep and
constrained by H and metal lines.
The similarity of the two measurements rules out ad hoc astrophysical
effects, such as the destruction of D without production of metals,
or unusual data problems. We are confident that we have measured
the primordial abundance of D
(log D/H $= -4.62 \pm 0.05$), which is the first such measurement because
other values in the literature are much larger, and apparently strongly
contaminated. The implied
baryon to  photon ratio does not agree with the values predicted from
measurements of $^4$He or $^7$Li. Since D is a simpler measurement
and simpler astrophysically, we propose that there are large systematic 
errors with the $^4$He data, and that $^7$Li is depleted by a factor of three
in halo stars. If this is correct, then
we have the first measurement of the primordial abundance of any
nucleon. We obtain a large value for the baryon density:
$\rho = 4.4 \pm 0.3 \times 10^{-31}$ g cm$^{-3}$ or
$\Omega_b = 0.024 \pm 0.002 h^{-2}$, such that 90\%
of baryons are unaccounted today. They are probably in the ionized 
intergalactic medium, and in the halos of galaxies, perhaps in some
cases as condensed MACHO objects seen in gravitational microlensing events.
}

\section{Significance of Deuterium} 

\subsection{Why Measure D/H}
The light nuclei deuterium (D or $^2$H), $^3$He, $^4$He and $^7$Li are 
created in big bang
nucleosynthesis (BBNS). In the standard model their relative abundances
depend on a single parameter: the ratio of the number density (cm$^{-3}$) of
baryons to photons $\eta $. The ratio of any two nuclei gives a measure
of $\eta $, 
but both D/H and $^3$He/$^4$He are favored ratios for the following reasons.
(1) These ratios are very sensitive to $\eta$.
(2) They use isotopes of the same element, which should have nearly
identical ionization, so that the observed ions are a good
measure of the total of each element: e.g. D~I/H~I $\simeq $ D/H,
where H~I is the astrophysical notation for neutral hydrogen.
(3) The abundance of D and $^3$He are relatively high, unlike $^7$Li.
Deuterium has three additional advantages over $^3$He:
(4) All observed D is believed to be produced in BBNS, and there is
complete destruction of D inside stars, so that observed D/H values
will be less than or equal to the primordial value. In contrast, stars 
both create and destroy $^3$He.
(5) The Lyman series lines of H and D lie at 912 -- 1216\AA\ , and are 
visible from space, or from the ground at redshifts $z>2$, but the
equivalent lines of He~II are at 228 -- 304\AA\ and are hard to 
observe, even at high redshift from space, because photons of these 
wavelengths are absorbed when they ionize H~I, which is very common
both in our Galaxy and in the outer regions of other galaxies. 
(6) The Lyman lines of D
are 82 \kms\, to the blue of those of H, where as the $^3$He lines are 
much closer to (13 \kms\, to the red) those of $^4$He.

\subsection{Astrophysical Uses of High Precision D/H Measurements}
There are six compelling reasons why we want to know the primordial
D/H to high accuracy.

{\bf The baryon density is a fundamental cosmological constant.}
The density of photons in the universe is accurately know from the temperature
of the cosmic microwave background, which has a Planck spectrum.
The measurement of primordial D/H should then give the density of baryons
at the time of BBNS, which is one of the few basic cosmological constants, like
the Hubble constant H$_0$, the total density of matter $\Omega_0$, and 
the age of the universe $t_0$. Baryons are one of the three most important
constituents of the universe, together with the cosmic microwave
background and non-baryonic dark matter.  The baryon density
plays a central role in the formation of the main structures in the universe: 
the intergalactic medium, galaxies, clusters, stars, as well as
the production of metals and the release of heat and ionizing radiation.

{\bf The cosmological baryon to photon ratio depends on high
energy physics, and may one day be calculated.}
The ratio of baryons to photons is set up prior to BBNS, possibly at
the weak scale (e.g. in a first order phase transition) \cite{ful94} 
or the GUT energy scales.
When we know the appropriate high energy model it 
may be possible calculate this ratio, and hence the observed value of the ratio
might be used to constrain parameters in the model, or to test models.

{\bf A precise value for $\Omega_b$ removes a degree of freedom from 
cosmological models.}
Prior to the measurement of D/H in QSOs, the ratios of the abundances
of the different elements allowed about a factor of five range in the
density of baryons. Our QSO data reduce the random error to 
12\%, a huge improvement. This type of precision is a powerful tool to
advance cosmology. When we learn the value of a basic parameter, we
remove a degree of freedom which previously helped models fit 
the data. We should get more powerful tests and more realistic models.

{\bf We can determine the fraction of baryons which are missing.}
Comparison of an accurate value for the baryon density at BBNS with 
counts of local baryons shows that $\sim $ 94\% are now unaccounted, enough
to account for dark matter in galaxy halos. Alternatively, if
the baryon density was low, few if any would be missing, and there would
not be enough for galaxy halos, which would have to be non-baryonic.  

{\bf Comparison of the primordial abundances of the light elements is a
powerful test of the physics in BBNS.}
The abundances of all light elements should be consistent with a single 
value of $\eta$, and the value of the test depends on the accuracy of the
abundances.

{\bf The changes in D/H with time help specify Galactic chemical evolution.}
The ratio of our D/H value, which is probably the primordial value,
with the value in the local ISM shows 
that $0.67 \pm 0.09$ of atoms in the ISM have not been in stars. This value
helps determine parameters which summarize galactic evolution, including
the mass locked up in stellar remnants and long lived stars \cite{edm94}.

\section{Deuterium in QSO Spectra}
Deuterium is a fragile nucleus which is readily converted into $^3$He and
heavier elements in
stars. The amount of D remaining in the local interstellar medium
depends on the history of chemical evolution. Edmunds (1994) has shown that
the fraction of primordial D remaining in the ISM should be $\geq  0.5$
 -- 0.7, which is the fraction of
(H) atoms in the ISM which have not been inside stars. This limit remains
true for arbitrary inflow and outflow, and rules out high primordial D/H,
but the precise numerical value
depends on the fraction of mass which is not returned to the ISM, and on 
the ratio of gas to total mass.

Deuterium in QSO absorption systems should be closer to the
primordial value, because most absorbers are in the far
outer parts of galaxies or in the intergalactic medium,
far away from most normal stars. Many absorbers are known to have low
metal abundances, 0.01 -- 0.001 of the solar value. This implies that
a negligible amount of D will have been destroyed, because the gas ejected by
stars which lacks D will contain metals. 

Adams (1976) \cite{ada76} 
discussed the two main requirements for the detection of Lyman
series absorption lines in QSO spectra. 
First, we need an absorber with a lot of H~I: a column density N(H~I)
\apgt $10^{17}$ neutral H atoms cm$^{-2}$, 
large enough that the D~I lines will be visible for D/H $\simeq 10^{-5}$.
These N(H~I) values are also large enough that the gas is partially or
completely optically thick to Lyman continuum radiation: photons with 
wavelengths $\lambda < 912$\AA\
(in the rest frame of the gas) ionize H~I to H~II, and are absorbed, producing a
drop in the QSO spectrum at $\lambda < 912$\AA . 
These absorbers are called Lyman Limit Systems \cite{tyt82}, and are readily 
identified in even low resolution spectra because they have a
sharp drop at the Lyman edge.
Second, the H~I in the absorber must have a very restricted range of 
velocities, to avoid doppler motions which would cover the D line, 82
\kms to the blue of H.
Most Lyman Limit systems do not show deuterium because the
H absorption is seen over a wide range of velocities, including those where
the D line would appear. Either the gas has too large a
velocity of dispersion (high temperature, and/or large turbulent motions), or
there are several absorbing gas clouds at slightly different velocities.
In most cases there is enough H to absorb
all the QSO flux at the expected position of D.

The line of sight to a QSO at $z \simeq 3$ intercepts about 6 absorbers
with sufficient H~I, but only about one of these will be at high enough
$z$ to shift the Lyman lines into the optical, where we can use 
large ground based telescopes.

Over the last 5 years we obtained medium resolution spectra of over 47 QSOs
with the Lick observatory 3-m telescope to search
for Lyman Limit systems with low dispersion velocities.
We applied three criteria to the Lick spectra to find
systems which might show D.
An ideal system should have a sharp Lyman limit, a \Lya line with a
small rest
frame equivalent width W(\Lya) $< 1$ \AA,
and weak metal lines \cite{tyt82}$^{,\,}$\cite{lan91}.
The sharp Lyman limit is only possible if there is residual flux between the
high order ($n \simeq 15$ -- 20) Lyman lines, which requires that those
lines are weak, with low equivalent widths. This requires that the H 
absorbs over a narrow range of velocities (low $T$, small turbulent velocities,
and absence gas clouds at different velocities) -- exactly what we need to
see D. The \Lya line must be weak if we are to see D
and weak metal lines are an indication that the gas comes from a restricted
range of velocities.
(We might occasionally see D in an absorption system which has a strong \Lya
line, but only if most of the H happens to be on the D side of the line, just
far enough away to avoid covering the D line.)

We have observed 15 candidate systems with the 10-m W. M. Keck
Telescope, and 2 have yielded deuterium measurements.
These 15 were selected from about 77 QSOs, including
30 with published medium resolution spectra.
The fraction of QSOs at $z \simeq 3$ which show D in optical spectra 
is about 3\%.

Several papers have presented upper limits on D/H in QSO spectra
\cite{son94}$^{,\,}$ \cite{car94}$^{,\,}$ \cite{rug96b}$^{,\,}$ \cite{wam96}, 
one paper gives a lower limit \cite{car96},
and one paper promotes an upper limit to a detection \cite{rug96}.
We now discuss our measurements on two QSOs, which are the only likely
detections, and then we discuss the other QSOs.

\section{D/H Towards QSO\,1937$-$1009}

Our detection of D in this QSO was described by Tytler, Fan \& Burles
\cite{tyt96}.  High quality Keck spectra
show strong absorption at the expected position of the D
\Lya line and weak but highly significant absorption at D \Lyb
at a redshift $z_{abs} =3.572$.
All the weak metal lines at this redshift, 
C~II, C~IV, Si~II, and Si~IV, appear to
have the same profiles, which are adequately described by two components
separated by 15 km~s$^{-1}$.
The velocity positions determined from the metals (Figure 1)
lie in the cores of
the higher order Lyman lines, and show the expected positions of the
deuterium absorption features.
The Lyman lines show that $>90$\% of the H~I must be in, near, or
between the two components, but not elsewhere because there are no other
lines in the spectrum which can explain the Lyman limit.
A simultaneous fit to the two components
gives a total column density log N(H~I) = 17.94 $\pm$ 0.05.
An additional component at $v \simeq +49$ \kms\,
produces S~IV, C~IV and H~I absorption, but it
is absent from the high order Lyman lines.
It has a low column density of log N(H~I) = 14.94 $\pm$ 0.14, and
it can not change D/H.
The total D in both components is log $N$(D~I) $= 13.30 \pm 0.04$, which is
fairly insensitive to the 
velocity dispersion and the precise velocity of the D
because the D lines are unsaturated.
We then have log D/H $= -4.64 \pm 0.06$, where this error includes
random photon noise, and fitting errors, but not systematic errors, which
are discussed in Tytler, Fan \& Burles \cite{tyt96}.

\begin{figure}   
\vspace{-0.7cm}
\centerline{\psfig{figure=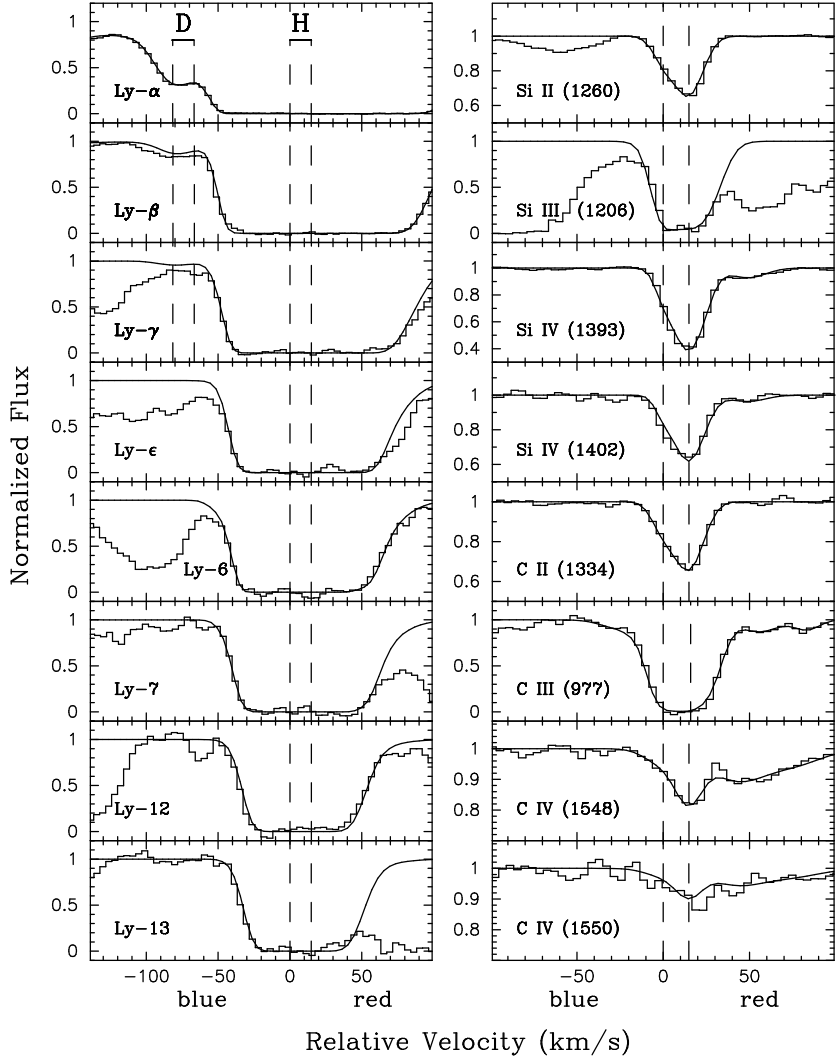,height=6.5in}}
\vspace*{-1.5cm}
\flushleft{
\caption{
Spectrum of absorption lines in QSO 1937--1009.
The Lyman series lines (left), and the metal lines
(right) arise in the same gas. The velocity scale is relative to
the redshift $z = 3.572201$ of the blue component (smaller wavelength).
The red component at $z = 3.572428$ is indicated by a second dashed line at
$+15$ km s$^{-1}$. 
The histogram represents the observed counts of the combined Keck spectra
in each pixel, normalized to the
quasar continuum.  The smooth curve shows the Voigt profiles convolved
with the instrumental resolution which produces the best
simultaneous fit to all the lines.
}}
\end{figure}

\subsection{Gas Temperature and Turbulent Velocity Dispersions}

Absorption line widths are determined by three factors. Instrumental resolution 
(known from calibration data to be 8 \kms\, FWHM), turbulent (bulk) motions,
and doppler broadening from thermal motions.
Keck spectra have sufficient signal to noise ratio (SNR) to separate these 
last two.
Turbulent motions effect ions of all elements equally, but
thermal broadening depends on the mass of the element, because of
equipartition of energy in the gas: light elements move faster.
We have determined the temperature $T$, and turbulent velocities
$b_{tur}$ of the gas which shows D \cite{tyt96}$^{,\,}$\cite{bur96}.

If Figure 2 we plot intrinsic velocity dispersion
(observed, corrected from instrumental resolution)
as a function of the mass of the ion, where
$b \equiv \sqrt{2}\sigma$ is the velocity dispersion.
The best fits for $T$ and $b_{tur}$ were determined without considering D.
Note that the $b$ values for D (the second data point from the left in each 
top panel) are very close to the expected value shown by
the curves. This must be the case if our model is
a good fit to the data, and the agreement provides additional evidence that
the lines are D and not some other ion.

The sound speed in the gas \cite{spi78}, $c = \sqrt{kT/\mu}$, given in Table
1, and are 10 -- 12 \kms, where 
the mean molecular mass $\mu = 1.22 \times $ mass of the proton. 
With thermal broadening, $b = \sqrt{2kT/m}$, thus we can equate
$b_{tur}$ to $\sqrt{2}c$, to find that the turbulence is subsonic and
quiescent, since the four clouds have $b_{tur} =$ 2.3, 3.2, 4.8 and 8.4 \kms.
If $b_{tur} > \sqrt{2}c$ the turbulence would have been be supersonic, and
we would expect an energy source which would set up shocks 
which create variations in the density and temperature.

\begin{figure}[h]
\vspace{-0.1cm}
\centerline{\psfig{figure=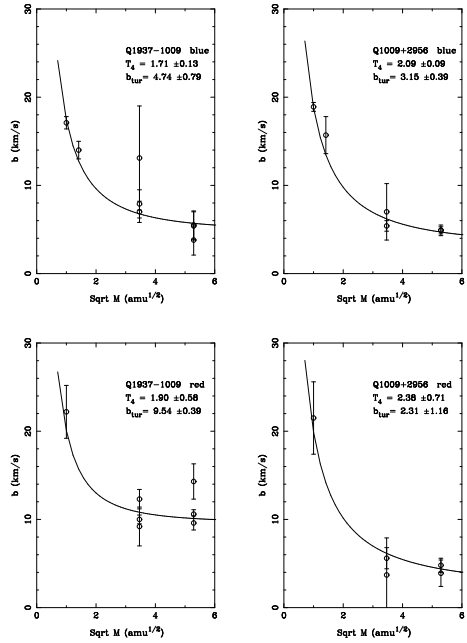,height=3.5in}}
\vspace*{-0.3cm}
\caption{The intrinsic velocity dispersion
(observed $b$, corrected for instrumental resolution)
as a function of the square root of the mass of the ion, in atomic
mass units. The ions shown are H, D, C and Si.
We do not show D in the red components (lower panels), because $b$(D) can
not be measured in those components, which are blended on both sides,
with the blue components of D and H.
The $b_{tur}$ is the asymptotic level of the curve running to the right,
where thermal broadening is negligible for high mass ions. The steepness
of the rise in $b$ to the left is a measure of the gas temperature, 
where $T_4 = T/10^{4}$~K.
}
\end{figure}

\subsection{The two gas clouds seen in QSO\,1937$-$1009}
The absorption system to \qn which shows D is
extremely unusual because it has only two main components, but it has
a large N(H~I) $\simeq 10^{18}$\cm2.
Typically we would see 5 -- 10 components in spectra with high SNR and
high spectral resolution, one or more of which would cover
up D, which is why D is hard to find.

The two components, which are a sufficient description of the spectral lines,
can have two different origins: a single gas cloud with a range of
metal abundance and asymmetric (non-Gaussian) turbulent velocity distribution, 
or two unrelated gas clouds.

First, with one gas cloud, the distribution of turbulent velocities would
mimic the $b$ values found in the two components fit to H.
If the different elements and ions are well mixed, then all absorption
lines would have the same velocity distribution, but
the spectra show that this is not true. In Figure 1 we see that
the D line is deepest in the blue component, but the metals are deepest in 
the red. The H~I is mostly (63\%) in the blue component, which contains only
$\simeq$ 20\% of the metals ions.
The metal abundances are about 7 times higher in the red component.
We can have either a gradient in abundances in a single, unmixed clouds, or 
two separate clouds. 
The sound crossing times in the clouds, $t_c = L/c$, are given in Table 1.
For the two clouds in \qn they are 0.6 and 1 Gyr, long enough that the
gas may not be mixed.
For the single cloud there could be two streams of gas mixing in the same
space. 

Second, there could be two gas clouds, which could be completely unrelated.
The physical separation corresponding to 15 \kms\, radial velocity could be
$ d_{sep} = v_{sep} / H(z) = 30 - 50 h^{-1} \rm kpc $, from Hubble
flow velocities alone, depending on the cosmology.  While the two components 
need not be physically associated, we speculate that they are tens of 
kpc apart in the outer halo of a galaxy. 

The small velocity difference between the two components is compatible with
gas in a galaxy halo. Velocity dispersions in halos today are hundreds of \kms, but much smaller velocities will be the rule for D, for two reasons.
First, absorption systems are so common that very few can come from the 
inner parts of galaxies. Relative velocities will be less in the outer parts
of halos, especially for gas which has not yet fallen in, which will be common
at high $z$. Second, relative velocities must be small for D/H systems 
($\sigma < 14$ \kms\,) because they
were selected to not have gas over a large range of velocities. Deuterium
could not be seen otherwise.

The difference in the abundances of the two components is not a surprise.
At early times, and in the outer regions of galaxies, there would be
a large dispersion in abundances in gas which had not mixed.

\begin{table*} 
\begin{center}
\caption{Properties of Gas which shows Deuterium.}
\begin{tabular}{|ccccccc|}
\hline
\hspace*{.3in}  & & & & & & \\
QSO & N(H~I) & $T_4^a$ & $b_{tur}$ & $2^{1/2}c$ & 
$L$ & $t_c$ \\
cloud   & (log)  & (K)      & (\kms)& (\kms) & (kpc) & 
(Gyr) \\
\hspace*{.3in}  & & & & & & \\
\hline 
& & & & & & \\
1937 blue & 17.74 & 1.62 & 4.8 & 14.4 & 11 & 1.1\\
1937 red~  & 17.50 & 2.36 & 8.4 & 17.4 & 7 &0.6\\
1009 blue & 17.36 & 2.1 & 3.2 & 16.4 & 5 & 0.4\\
1009 red~  & 16.78 & 2.4 & 2.3 & 17.5 & 1 & 0.08\\
& & & & & & \\
\hline
\hline
\end{tabular}
\noindent{$^a$}{$T_4 = T/10^4$~K. Values and their errors are derived in
Tytler, Fan \& Burles (1996) and Burles \& Tytler (1996)
}
\end{center}
\end{table*}

\subsection{Why the Data on QSO\,1937$-$1009 do not allow a High D/H}
We would need to increase N(D), or decrease N(H), neither of which is possible.
If we increase N(D) we would see dramatically more absorption. The D line is 
unsaturated, absorbs about 0.7 of the flux in its core, and is seen in 
data with SNR 75 per pixel of 4 \kms . 
It extends over $>10$ pixels in \Lya and it is also seen in \Lyb.
The 1 $\sigma$ uncertainty in N(D) is 10\% (\cite{tyt96}), so a
ten fold increase is ruled out at the 70 $\sigma$ level.
Adding contaminating H to the D line acts in the wrong direction;
there is then less D, and a lower D/H.

The situation with H is as tight \cite{tyt96}.
The 1 $\sigma$ uncertainty in N(H) is only 12\%, which is
precise for several reasons which
do not apply to most other absorption systems which have
been analysed in the past. It is incorrect to generalize that
we can not obtain accurate N(H~I) whenever H lines are saturated.
(1) We have high SNR and high resolution.
(2) We have done a simultaneous fit to 13 H lines and the Lyman continuum.
The oscillator strength of Lyman-19 is 2000 times smaller than that
for \Lya which means that their absorption lines profiles are
significantly different, which gives sensitivity to N(H~I) even when
lines are saturated.  Lyman lines 12 --  18 are clearly seen and resolved,
and give 
log N(H~I) $= 17.7 \pm 0.07$, with a $2 \sigma$ limit
at 17.94, which is the best fit to all the Lyman lines.
(3) The Lyman continuum absorption also requires
a high log N(H~I) $> 17.8$. We have looked for other absorption systems in
the spectrum which might account for this Lyman continuum absorption.
There are none, and all of the absorption comes from the system which shows
D. 
(4) 
Adding a third or fourth component will not change the answer significantly,
as long as their redshifts coincide with the metal lines.

We know that the continuum near 4180 \AA\ is well determined because 
the fit to the Lyman  lines is good,
and there is little absorption between the Ly-series.
The  SNR weighted flux in  the region 4125 - 4175 \AA\ is $0.015 \pm 0.014$, 
where the continuum level is 1.0, which corresponds to an optical depth
of $\tau > 3.5$ ($1 \sigma $) and log N(H~I)$ > 17.74$.

\begin{figure}[h]
\vspace{-0.1cm}
\centerline{\psfig{figure=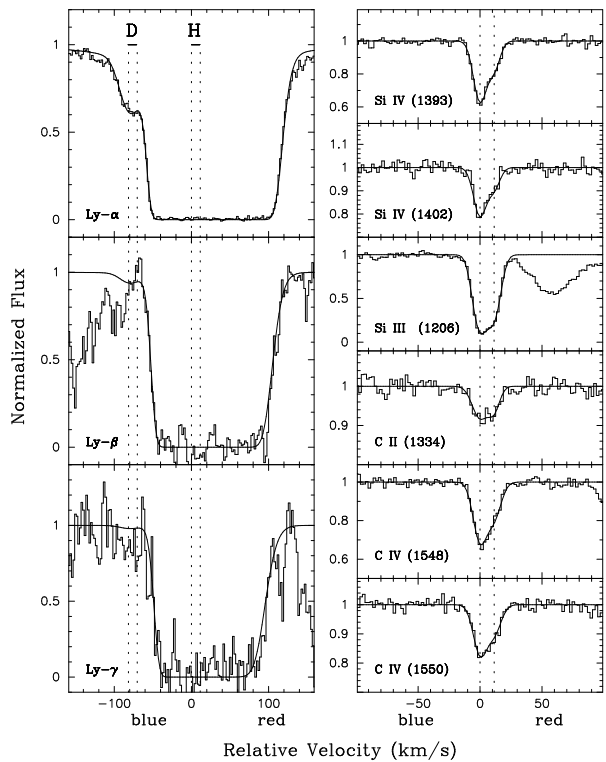,height=4.5in}}
\vspace*{-1.0cm}
\caption{
As Figure 1 but for \q2 (emission redshift $z_{em}=2.616$, V=16). 
Keck spectra of Lyman $\alpha, \beta, \gamma$ (left), and
all the metal lines (right) in the absorption system.
Zero velocity corresponds to the redshift $z = 2.503571$ of the blue component.
The red component at $z = 2.503704$ is indicated by a second dashed line at
$+11$ km s$^{-1}$.
}
\end{figure}

Models which give a higher D/H by lowering N(H~I) must leave
residual flux below the Lyman limit, unless the H is put in other absorption
systems. 
We stated in Tytler, Fan \& Burles (1st sentence in 4th paragraph)
that all the H must lie in the main absorption
system, because the Keck spectrum of the \Lya forest shows that 
there are no other absorptions systems with large N(H~I) in this region.
To the blue of the D/H absorption system (z = 3.572),
the only moderate column density systems are ($z=3.555650$, log N(H~I) =
15.06, $b=$ 27), ($z=3.553659$, log N(H~I) = 15.11, $b$=53.8),
and ($z=3.256067$, log N(H~I) = 15.70, $b$= 36.0).
These systems do not have enough N(H~I) to have a noticeable effect 
on the flux blueward of 4180 \AA\, hence all of the H~I which causes the Lyman 
break is in the D/H system.

We conclude that there is no possibility that D/H is much larger than we
stated.  Continuum errors could systematically increase log D/H by 0.06
(15\%), but not much more.

\section{D/H Towards \q2}

In Figure 3 we show nine lines in the absorption system at $z=2.504$
towards QSO 1009+2956.  Details are given in Burles \& Tytler \cite{bur96}.
The residual flux below the Lyman edge gives an accurate
and independent measure of
all H~I in this velocity region \cite{bur96}, log N(H~I) = 17.46 cm$^{-2}$.
There are no other \Lya lines within 5000 \kms\, of $z=2.50$ which
have H~I column densities $>10^{16}$ cm$^{-2}$, so that all of the
Lyman continuum absorption must be produced by gas in the $z=2.504$ absorption
system.
The blue side of the \Lya, \Lyb and \Lyg lines are best fit if all
of this H~I is near the two velocity components which are seen in the metals.
There could be be additional H at velocities between the metal lines and
$+40$ \kms, provided this gas has very low metal abundances [C/H] $< -3.5$.
But, nearly all known QAS with large N(H~I) have metal
abundances [C/H] $> -3$.
A simultaneous fit to the Lyman $\alpha$, $\beta$, and
$\gamma$ lines in the Keck spectrum and the Lyman continuum
absorption in the Lick spectrum (Figure 3)
gives D/H = $3.0 \, ^{+0.6}_{-0.5} \times 10^{-5}$ ($1\sigma $ random 
photon and fitting errors).

\subsection{The two gas clouds seen in \q2}
Unlike QSO\,1937$-$1009, for \q2 the metal and D lines have the same
velocity profiles, so that the two velocity components have the same
metal abundances.  This rules out mechanisms which
might separate metals from D and H.
For this QSO it is more likely that the two components could
be part of the same gas cloud, which has an asymmetrical distribution of
turbulent velocities.
The blue component has 80\% of both the H and metals.

\subsection{Why the Data on \q2 do not allow a High D/H}
The random error on N(H~I) is 12\%, while that on N(D~I) is 15\%, both
very small. The $1\sigma $ error on D/H is $0.5 \times 10^{-5}$, so that
D/H $\simeq 25 \times 10^{-5}$ is ruled out at the $50\sigma$ level.
A larger D/H would require a much lower N(H~I), which is ruled out by
the Lyman continuum absorption. Again, there are no other \Lya lines
in the spectrum which can account for this absorption.
We conclude that D/H must be low in this absorber.

\section{Detections -- Not Upper Limits}

A major difficulty with the measurement of D/H is that there is
additional H absorption at a large fraction of wavelengths where D
could be seen, which is always in the Lyman alpha forest.
This additional H can blend with real D lines, it can completely swamp the D
line so that the signal is mostly H, it can produce a saturated line
with no remaining flux in the core,
and it can change the H lines which are associated with the D.

For our two QSOs the main effect will be slight increases in the amount of 
absorption at the position of D. There will not be any noticeable change
to N(H~I) because we see many H lines, we reject those which are blends
(absorbed more than expected) and we see Lyman continuum absorption.

We used Monte Carlo simulations of spectra \cite{tyt96} to estimates
the amount of H absorption on top of the D, and we have corrected
our D/H values for this additional H. The corrected D/H values are then
measurements and not upper limits.

In Figure 4 we show the log likelihood as a function of D/H.
For each D/H we made a template spectrum showing the D and H
\Lya lines, where all parameters are fixed by the $T$, $b_{tur}$ and
$z$ values 
given in Table 1. We made 500,000 template spectra for each D/H, and to
each we added different \Lya forest lines selected at random from known
distributions of N(H~I) and $b$
(a Gaussian centered at b=28 with 
$\sigma = 6$ \kms\, and no values with $b\leq 12.5$).
The likelihood is calculated comparing each template with the Keck spectrum.
For \qn the most likely D/H is only slightly lower that the value
obtained by fitting the data, but for \q2 the most likely D/H is lower than
the fit value by about $1\sigma$. 

The likelihood plots show that there is a negligible chance that the D line
in \qn is significantly contaminated by H from the \Lya forest.
There are three reasons why a combination of one or more H~I lines can not
reproduce even a small portion of the column density at the position of D.
First the D line is deep -- it absorbs 70\% of the flux in its core --
and it would require lines with large N(H~I) to match this absorption.
Such lines are rare.
Second, the line has a very steep blue edge, which requires
a low $b$ value.  Even if D/H in the two components is at the ISM value,
and additional H at some near by velocity
accounts for most of the absorption where D is expected, we find 
this contaminating H would have $b<18$ \kms, which is low enough to be unusual.
Very few H~I lines are this narrow, especially at high N(H~I).
Third, the SNR is high, 75 per 4 \kms pixel, and D extends over $>10$ pixels, 
which means that the fit must be extremely good to get a large
likelihood that the data comes from the model.
 
But for \q2 the likelihood plots show that we should expect
some H on top of the D. This is because this D line is not as deep 
(it absorbs 0.4 of the flux in its core), and it is not as steep 
as the D line in \qn: the errors now permit a larger $b$: $15.7 \pm 2.1$
\kms\, versus $14.0 \pm 1.0$.

The chance of contamination will be much larger in data with lower SNR or
lower spectral resolution, or when fits are poor or non-unique,
because we are then less able to distinguish
between a D profile, and a blend of H lines which appear similar.
Lines with large $b$ (similar to \Lya forest lines), and low N(H~I),
(most forest lines have low N(H~I)), are also most likely to be 
contaminated with H.

\begin{figure}[h]
\vspace{-0.5cm}
\centerline{\psfig{figure=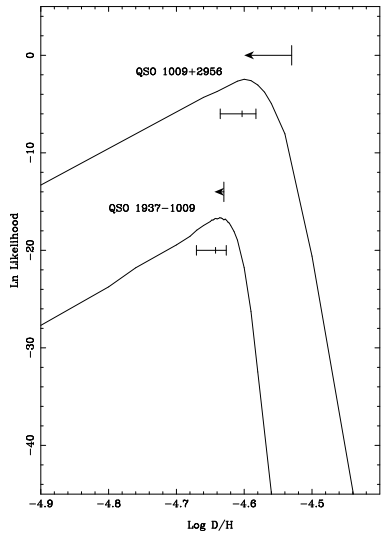,height=3.0in}}
\vspace{-0.0cm}
\caption{Natural Log Likelihood as a function of D/H in 
our two absorption systems.  The arrows show the shift in the expected
value of D/H when contamination from weak \Lya lines is removed.
The error bars show the means and standard deviations of the likelihood
functions.  
The normalizations of the likelihood functions are arbitrary.
}
\end{figure}

\subsection{D/H could be Lower: Blending with H in the Lyman limit system}
We have described the statistical correction of D/H
for unrelated \Lya forest
lines at the location of D. This simulation does not correct for
H which is statistically or physically associated with the gas
which showed the D. Such H could lie on top of the D line, and increase
the measured D/H above the true value.
It would not effect N(H~I) unless its column density was comparable
to the main components, in which case it would have been detected.

We can not simulate such correlated gas, because we do not how it is
distributed.
We need the distribution in velocity  (two-point correlation 
function) on scales down to 40 \kms, and the distribution in
$T$, $b_{tur}$ and metal abundances (metals would be seen if present --
ruling out contamination of D)
of all H~I in absorption system complexes with total N(H~I) comparable to
those which show D.
This data will be difficult to obtain because, (1) the H lines
are normally strongly blended, and (2) we will need to observe tens of
systems to see if the distributions are same for all types of systems,
e.g. those with different numbers of components, or different total N(H~I).
This is important because the gas which shows D is not a random sample of
Lyman limit systems. They were pre-selected to be suitable for the 
detection of D:
steep Lyman breaks, weak H and weak metal lines, which together imply
few velocity components, low temperatures, and low $b_{tur}$.

If there are differences, then we may have to limit the sample to
those systems which show D, which means that we will have many D/H
measurements before we can make corrections for associated H.

We can use the red side of an absorption system 
to determine the properties of the associated H, because 
the red and blue sides in many different gas clouds
will be statistically identical except for the D in the blue.
For both our systems there is one additional low N(H~I) absorbing component on  
the red. For \qn\, this extra component has $b=33.5$ \kms\, at $v=49$ \kms, 
and for \q2\, the extra component has $b=31$ \kms\, 
at $v=40$ \kms. These two cases show 
that the associated gas does not necessarily have a low $b$ value, and in 
neither case could this gas look like or contribute significantly to the D 
lines, which have much smaller $b(D) \simeq 15$ \kms.

\subsection{What if the two clouds have different D/H?}
For \qn\, and \q2 we could not measure the $z$, N(H~I) or $b$ values for 
D in the red components because they were blended on both sides, with
the blue components of H and D. 
Some absorption is required here to account for the spectrum, so we set
(D/H)$_{red} =$ (D/H)$_{blue}$.
If we relax this constraint, for \q2 we find for the blue
component: N(H~I) = $17.3 \pm 0.9$,  N(D~I) $ = 12.94 \pm 0.05$ and
                 log D/H $= -4.4 \pm 0.9$. Here N(H~I) has gone down
15\%, N(D) up 26\% and D/H up 32\%. But the random error on D/H is now
a factor of 8, rather than 20\%.
For the red component N(H~I) $= 16.6 \pm 8.2$, N(D~I) $= 12.0 \pm 0.3$,
and log D/H $= -4.6 \pm 8.2$. Huge errors.
We get much smaller errors when we add the constraint that the
total H and D must together match the spectrum. One way to introduce this
constraint is to take D/H down to the ISM value in the red component,
and conserve the total D by increasing D in the blue component.
For \qn\, we get (D/H)$_{blue}$ =$-4.57$, which is still a low value.
A second way is to tie to have the same D/H, as we did in our analysis
\cite{tyt96}$^{,\,}$\cite{bur96}.

\section{Size and Mass of the Absorbing Clouds}

\subsection{Length down the line of sight}
We can calculate the length of the absorbing gas by
dividing the column density of neutral hydrogen by its estimated number density,
$L = N(H~I)/n_{H~I}$.  The column densities are well specified by the
optical depths at the redshifts of the absorption systems.  In our
two systems we estimated the neutral fraction of hydrogen 
$n_{H~I}/n_H = 10^{-2.5}$.  The number density of hydrogen is calculated
from its relation to the ionization parameter, $U=n_\gamma/n_H$, 
where $n_\gamma$ 
is the number density of photons with energies above one Rydberg.  Using
a background mean intensity of $J_\nu = 10^{-21}$ ergs cm$^{-2}$ s$^{-1}$
Hz$^{-1}$ sr$^{-1}$, and ionization parameter of $U = 10^{-3}$, we find
$n_H \approx 0.005$ cm$^{-3}$.  For QSO\,1937$-$1009 the blue component with
log N(H~I) = 17.74 has $L = 11$ kpc, and the red component has
log N(H~I) = 17.50 and $L = 7$ kpc.
For \q2 the blue component has log N(H~I) = 17.36, $L = 5$ kpc and the
red component has log N(H~I) = 16.78 and $L = 1$ kpc.  A characteristic 
thickness (along the line of sight)
of our Lyman Limit systems is $L = 10$ kpc.

\subsection{Mass probed by the quasar light}
The total mass depends on the shape of the gas cloud.
The mass in the line of sight is proportional to the volume of the cloud, 
$M = \rho \, \pi \, r^2 \, L$,
where $\rho = \mu \, n_H \approx 10^{-26}$ g cm$^{-3}$ 
is the mass density of the gas, 
and $r$ is the radius of the background light at 
the redshift of the absorption system.
In both of our systems, emission redshifts $z_{em} - z_{abs} \ll 1$, 
so $r$ is approximately the size
of the optical radius of the QSO.  We assume this radius to be $r = 1$ pc,
and find $M \approx 5 M_\odot$, for a thickness of $L = 10$ kpc.

This mass is a tiny fraction of the entire absorbing cloud, 
since the quasar light
probes a very narrow ``pencil-beam" down the line of sight.  
Since absorption lines
measure the total optical depth in this pencil-beam, 
any localized destruction
of D will be insignificant to the total optical depth. This rules out
destruction of D in stellar winds \cite{rug96b}.

The total mass depends on the shape of the absorbing cloud. If they are
spherical $M \approx 2 \times 10^9M_\odot$ for a diameter of $L = 10$ kpc.
If they are flat, then we will see more of those which are near face on,
because they present a larger cross section, and because the D/H systems
are biased to have simple velocity structure down the line of sight.
The total masses of observed clouds will then average larger than the masses
estimated assuming spherical shapes.

\section{Destruction of D in stars}
The ISM D/H is $0.67 \pm 0.09$
of the value which we measure. This implies that 0.33 of D is destroyed when
the Oxygen abundance rises to about 0.3 of the solar value.
If the destruction of D is proportional to metal abundance, then
$<0.002$ of D would be destroyed when the O abundance is $< 0.003$ of solar.
This is conservative because the present ISM involves stars with a variety
of masses. At early epochs destruction would mostly be by higher mass stars,
(because low mass stars are still on the main sequence and have not ejected
much matter)
which eject more metals relative to H than do low mass stars. This gives
less D destruction per metal atom added to the gas.

\section{Comparison with other D/H Possible Detections in QSO Spectra}

Table 2 lists the six absorption systems which might show D,
along with a recent measurement of D in the ISM.
Notice the wide range of D/H values.
How can a single value for primordial D/H arise from values spanning
a whole order of magnitude?  

The values of D/H in the absorption systems should not be weighted equally.
Our two values of D/H have statistical errors of 20\% or less, while
the other detections have errors greater than 50\%.  In regards to absolute
sensitivity to D/H, our two measurements are 10 times more sensitive than
the others.  Ours are sensitive to D/H at 1 part in 10$^5$, 
while the others are only
sensitive to D/H in parts to 10$^4$.  We now compare the other measurements
of D/H listed in Table 2 to our measurements.

\begin{table*}
\begin{center}
\caption{Deuterium in QSO Spectra.}
\begin{tabular}{|cccccc|}
\hline 
\hspace*{.2in}   & & & & & \\
QSO & $z_{abs}$ & D/H $^a$ &
N(H~I) $^b$ & [C/H]$^c$ & Reference \\
&           & ($10^{-5}$) & (log) & & \\
\hspace*{.3in}   & & & & & \\
\hline
& & & & & \\
ISM     & 0.0   & 1.6 $\pm$ 0.1 & 18.2 & ... & \cite{lin95} \\
(A) 1009+2956 & 2.504 & 2.5 $\pm$ 0.5 & 17.46 & $-$2.9 & ... \\
(B) 1937$-$1009 & 3.572 & 2.3 $\pm$ 0.3 & 17.94 & $-$2.2, $-$3.0 & ... \\
(C) 0014+8118 & 3.320 & $\leq 25$ & 16.7 & $<-$3.5 & \cite{son94}$^,$
\cite{car94}$^,$\cite{rug96} 
\\
(D) 0014+8118 & 2.798 & 19 $\pm$ 9 & 18.04 & $-$2.5 & \cite{rug96b} \\
(E) 1202$-$0725 & 4.672 & $\leq 15$ & 16.7 & ... & \cite{wam96} \\
(F) 0420$-$3851 & 3.086 & $\geq 2$ & $\geq 18$ & $-$1.0 & \cite{car96} \\
(G) 1422+2903 & 3.515 & 200 $\pm$ 70 & 15.25 & ... & \\ 
\hline
\end{tabular}
\end{center}

\noindent{$^a$}{We show the approximate $1\sigma $ random error.}

\noindent{$^b$}{Column density in logarithmic units of cm$^{-2}$.}

\noindent{$^c$}{Carbon to hydrogen ratio in logarithmic units, 
relative to solar.}
\end{table*}

{\bf Q0014+8118, $z_{abs} = 3.320$:} 
The QAS towards Q0014+8118 differs from our two in several ways.
Since no metal lines were detected,
the velocity structure of the cloud could be determined only by
median filtering of the higher-order lines in the \Lya forest \cite{son94},
 and the Lyman-$\alpha$
feature was considerably more complex: five components were required
for an adequate fit (\cite{car94}).
The neutral hydrogen column density of
the component where deuterium is measured, log N(H~I) = 16.74, is
4 and 10 times lower than the column densities in our two
QAS, which reduces the sensitivity to low D/H.
The published fits to the D line are either poor \cite{rug96}, or require 
additional H on both sides of the line \cite{car94}.

{\bf Q0014+8118, $z_{abs} = 2.798$:} Rugers \& Hogan (1996b) identify a 
second deuterium ``feature"
in the same Keck HIRES spectra.  But the identification is not a feature,
only the blue wing of a \Lya absorption complex.  They model 
deuterium \Lya as saturated, giving a column density of log (D~I) = 14.31
$\pm$ 0.25.  This absorption will be hard to confirm because
\Lyb is to the blue of the Lyman edge of the other absorber, where there is
little flux.  With only the wing of \Lya
in low S/N spectra (S/N $\approx$ 20), and no higher order Lyman lines,
we conclude that this identification of deuterium is far from certain.

{\bf Q1202-0725, $z_{abs} = 4.672$:} Although the high redshift of this
system should allow a measurement of D at earlier epochs, the density of
\Lya lines significantly increases the chance of interlopers.  In addition,
this system may have a very high oxygen abundance, twice that of
the sun \cite{wam96}.
The column density of N(H~I) is poorly constrained, due to the
low optical depth at the Lyman continuum and the contamination of higher order
Lyman lines by the \Lya forest. 
The spectra have relatively low SNR, and the fit to D is not tightly
constrained.
We conclude that this system is not well suited for a measurement
of D/H.

{\bf Q0420-3851, $z_{abs} = 3.086$:} This absorption system is very complex,
over 9 components are required for an adequate fit.  The lower S/N $\approx$ 10
spectra do not tightly constrain H~I or D~I.  The metallicity is also
high, approximately 1/10 of solar.   The confusion associated with this system
makes this a poor candidate for an accurate measurement of D/H. 

\section{What is the Cosmological D/H Value?}
There are only three claims of a detection of D: our two and that by
\cite{son94}. All others were first presented as upper limits, as
was that by \cite{son94}. Our two D/H values agree, while the third is
10 times larger.  There are three options:

1. The cosmological D/H is low, at the value which we see in two QSOs.
The high values reported by others are all contaminated by H lines,
and in many cases the SNR is too low to see D even if there
was no contamination. We show below that this explanation provides a complete 
description of all D/H data.

2. The cosmological D/H is high, at the value suggested by \cite{son94}.
Our data disagree with this high D/H at the 50 and 70 $\sigma$ level,
so D must have been destroyed by some unknown astrophysical process.

Rugers \& Hogan (1996) speculate that winds from high mass stars might
eject gas which lacks D and metals. This process is ruled out for
three reasons. First, the absorbing clouds are about 10 kpc along the line of 
sight, but winds from individual stars will travel pc. We would need
many stars, lined up along the line of sight.
Second, these stars will explode and eject metals a short time after their
winds eject H without D. We would need to synchronize the star formation
along the 10 kpc line of sight, so that nearly all stars were in the stellar
wind phase at the time of observation. And we would need  to avoid
high mass stars which explode shortly after they form.
Third, we see identical D/H in two lines of sight. Towards QSO\,1937$-$1009
we must remove 90\% of the D, while towards \q2 we must remove
87\%. Hogan agrees (private communication, May 1996) that this rules out their 
suggestion. 

Hypothetical mechanisms which destroy D without making metals must also
meet other constraints. They must
not operate in the local interstellar medium (ISM), where D/H is 
approximately constant, and they must not take D/H to values lower than 
in the ISM, and they not change the C/Si abundance ratios 
which we measure to be the same as in Galactic halo stars.

For these reasons, we also disagree with Schramm \& Turner \cite{sch96}
who stated ``it may be that the cloud is very inhomogeneous and the heavy 
elements have been dispersed. This is not implausible, because the material 
giving rise to the absorption lines has only a tiny fraction of the cloud's 
mass." The gas in QSO\,1937$-$1009 is clearly inhomogeneous (the two clouds
differ in $z$, metal abundance, $T$, and $b_{tur}$), but we do not see
how metals can be removed from two lines of sight, each 10 kpc long, leaving
the same D/H in each. The turbulence along both line of sight is subsonic, 
with very low bulk motions ($b_{tur} <2 $ -- 8 \kms), and  no sign of shocks.
And where are the metals which have been ejected from the gas clouds?
They should be seen.
These absorption systems have high N(H~I) and low metal abundances.
If the metals were ejected, they should create high metal abundances
in gas with low N(H~I) (high N(H~I) is very rare), which is not common.
We would also expect that the metal, H and D lines would have different
velocity profiles, but they are similar, especially for \q2. 

The observation of the same D/H in two lines of sight
should rule out all unusual astrophysical effects, and all chance data glitches.

3. The universe is inhomogeneous and D/H is high in some places, and 
low in others. The natural mass scales for inhomogeneities
in isocurvature fluctuations are compatible with the observations:
$10^6$ -- $10^{11}$ solar masses \cite{jed95}.
But the mean cosmological D/H should be about $7 \times 10^{-5}$, higher
than suggested by our two measurements. And we would 
expect to see a variety of D/H values, such that it might be hard
to understand two values which agree to within 10\% and a third which
differs by a factor of ten. However more high quality
measurements will be needed to determine if D/H varies spatially.

\section{False Identifications}
In Figure 5 we show Keck spectra of most of the \Lya forest of QSO 1946+7658.
Note the extremely large number of lines, and especially the saturated
lines  which go to zero flux in their cores, as must all with enough N(H~I) to
show D. Any of these saturated lines might have enough N(H~I) to show D;
it is hard to tell because we must fit each line to
find its N(H~I). All saturated lines have log N(H~I) $\geq $ 14.5, 
unless they happen to be a close blend of lines with lower N(H~I).
But most saturated lines
have 0.001 -- 0.01 of the N(H~I) needed to show D.
Lines suitable for D must have both high N(H~I) and low $b$, so they are
not necessarily the widest lines.
This spectrum shows why we prefer to see the Lyman limit and metal lines, 
before we make a claim to a detection of D.

\begin{figure}
\vspace*{-1.3cm}
\centerline{\psfig{figure=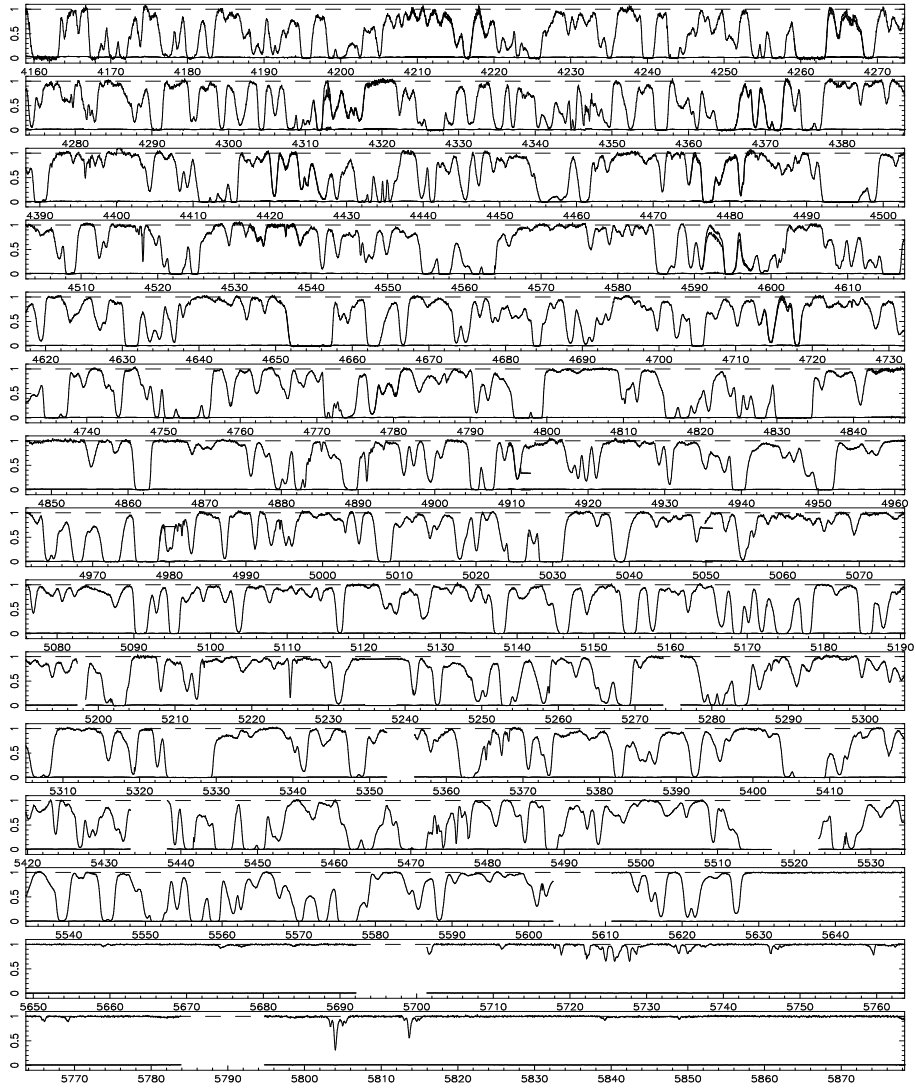,height=7.0in,width=5.8in }}
\vspace*{0.5cm}
\vspace*{-2.2cm}
\flushleft
\caption{
Keck HIRES spectrum of QSO 1946+7658 at 8 \kms\, resolution and SNR $>100$
in many pixels. Most of the absorption lines are \Lya forest lines.
Lines which have zero flux in their centers are saturated, and about one of
these lines might have enough N(H~I) to show D, if it also had a low
velocity dispersion. The last two panels cover wavelengths 
outside the \Lya forest, where there are no H~I lines,
because $\lambda > 1216 \times z_{em}$.
}
\end{figure}

In Figure 6 we show a portion of the spectrum of one QSO which appears like
D next to H. We spent 1 hour searching one spectrum to find this
example. The fit with one component for D is poor, and should be rejected.
The fit with two components (Figure 7) looks excellent, and gives
D/H = $2.1 \pm 0.7 \times 10^{-3}$. We do not think this is D, because
such chance occurrences of a weak H line just to the blue of a strong
saturated line are common, and examples can be seen in Figure 5.
This explains why Rugan \& Hogan are able to find two examples of D/H
in the Keck spectrum of one QSO, where as we believe that D is seen in about
3\% of QSOs.

\begin{figure}[h]
\vspace*{-0.6cm}
\centerline{
\psfig{figure=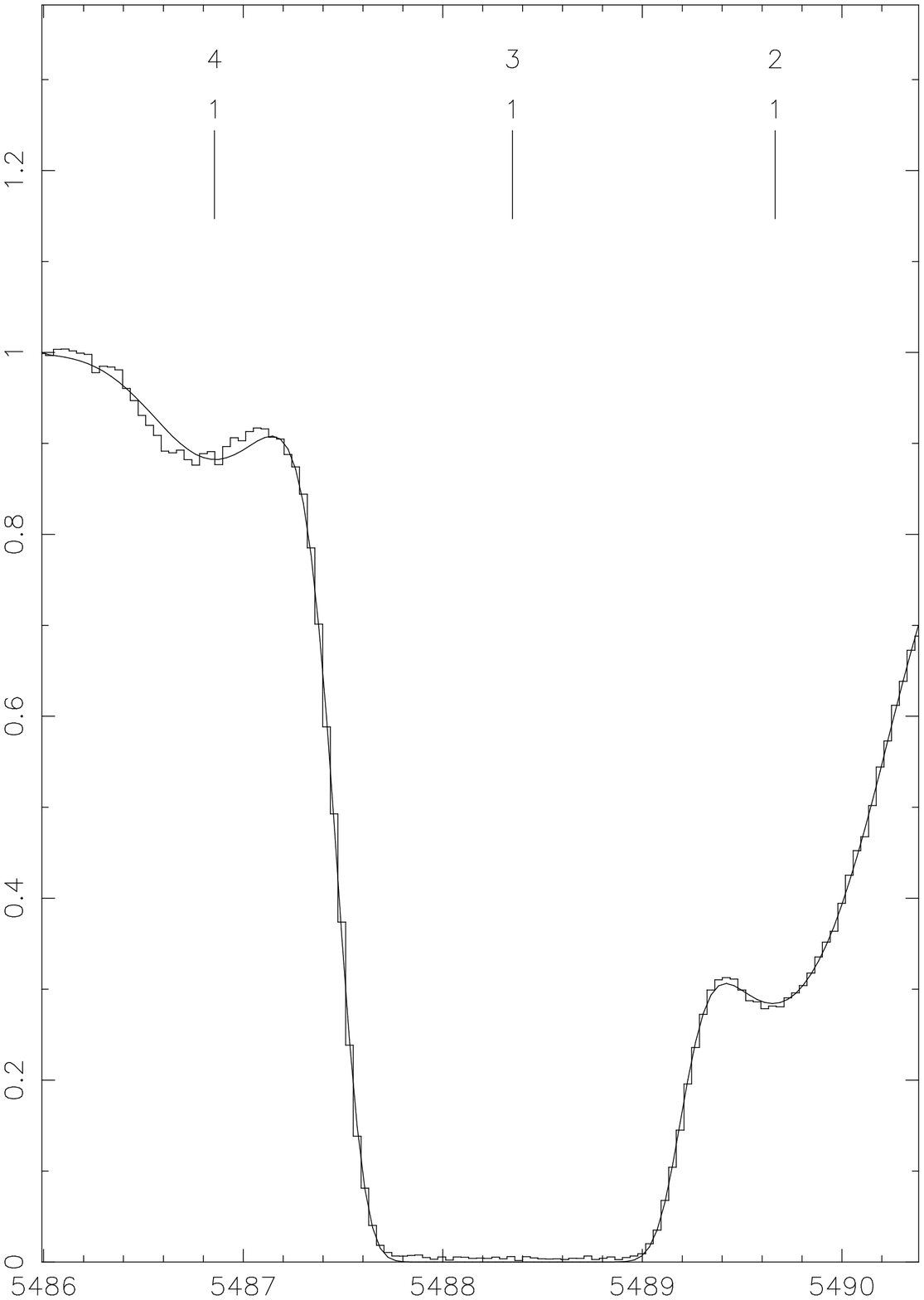,width=2.0in}
\psfig{figure=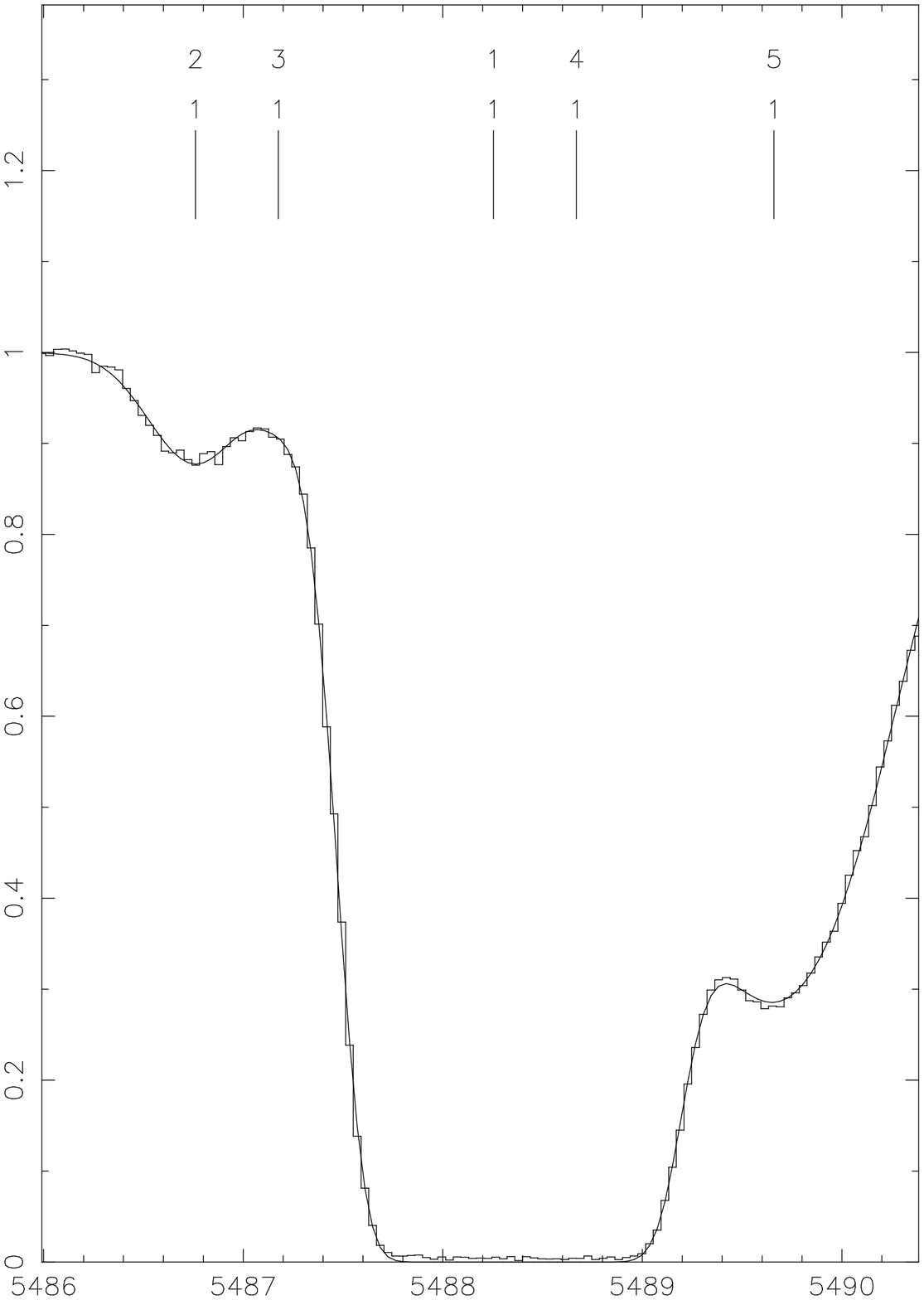,width=2.0in}
}
\vspace*{-0.2cm}
\caption{
Keck HIRES spectrum of QSO 1422+2309 with SNR 100 per pixel.
The fit on the left used only one H component to fit the saturated 
absorption line and one component to fit the correspsonding D line.
The fit is poor, and rejected.  
The fit on the right has two critical components for H (11 and 41)
and D (21, 31).
This fit looks much better, and gives an extremely large D/H, but we
do not accept it as a D/H measurement, because
there is a large chance that the D line is H, for the following reasons,
which are explained in the text.
(1) There is no supporting evidence [metal lines for $z$ values,
predictions for $b$(D), estimates of temperature, or element abundances
which would indicate that the fit is reasonable].
(2) We required two components to fit D. These were determined {\em ad hoc}
to get a fit to D and H, without additional evidence.
(3) N(H~I) is low.
(4) We searched only one QSO for this example.
(5) The fit is actually poor with a reduced $\chi ^2 = 4.1$ per degree of
freedom over 114 pixels, which is
hard to see because the SNR is so high.
We reject this as D/H even though there are some favorable characteristics:
(6) The SNR is very high, 100 per pixel.
(7) The spectral resolution is excellent, 8 \kms.
(8) The D line is clearly resolved from H.
(9) There is no doubt that there is a feature at the position of D.
In these respects this D/H candidate is better than most of those in
the literature.
}
\end{figure}

\subsection{Fits should be excellent and unique}

The higher the SNR of the data, 
the harder it is to get an acceptable fit,
because we see the details of the velocity distribution of the absorbing gas.
Fits must be excellent in the regions of the spectrum where one fit is
distinguished from another with very different D/H.
It is not sufficient that a fit have a reasonable $\chi ^2$ over the whole line.
The $\chi ^2$ must be good over just those pixels which are 
critical to the fit. Inclusion of additional pixels increases the number 
of degrees of freedom, and can hide a fit which is poor in the critical regions.

Two of the published fits to QSO 0014+8118 are unacceptable, while the third
is not unique because it has too many free parameters.
Songaila et al.\cite{son94} 
show a D line which is much too narrow to
account for observed absorption  (their Fig 3). 
Additional absorption, probably H,
is required on both sides of the line, as was shown by Carswell et al.
\cite{car94} (their
Fig 2).  This fit is good, but it is not unique because
they must introduce 6 new free parameters (3 per new H line) to fit the
region which could contain D. There extra parameters greatly increase 
the chance that the line is almost entirely H. 

Rugers \& Hogan\cite{rug96} reanalyzed the published spectra of Q0014+8118,
and determined that the deuterium feature is better
fit with two very narrow components separated by 21 \kms.
Their model is unphysical, because it absorbs flux at wavelengths where 
flux is seen: e.g. in the Lyman lines at 
3974, 3972, 3962.5, 3961, 3959.5, and 3954.8 \AA\ (their fig 2).
The fit is also poor in other places: D~I \Lya, H~I \Lyg and \Lyd blue
wing and central spike. It is especially unfortunate that the fit is bad
right in the center of D at 5251 \AA , 
where the data shows a spike where there model specifies
there should be none (their fig 1).
This is the critical part of the spectrum, which determines if there is one or
two components to D. It seems that these components were chosen ad hoc
to fit the
hypothesis that the line was D: they do not fit the data, and they are not
motivated by other lines. For our two QSOs the two components are
required by metal lines.

The Rugers \& Hogan\cite{rug96b} fit to a second D/H candidate in the same 
QSO (0014+8118)
is not acceptable because it is not unique. Here the blue edge of a
completely saturated line is assigned to D, while the rest of the line
is composed of 7 additional components which were seen in metal lines.
It seems more likely that the D line is another H line. There is no reason to
suppose that this is D. 

\subsection{Ideal Systems}

The chances of false identification of a pair of H lines as D and H is
reduced in the following circumstances.

\subsubsection{Search Many QSOs} 
It is best to search many QSOs and concentrate on the best few systems.
The chance of a false identification increases greatly when a lot of
attention is spent looking for any features which might be D/H in
a few spectra. We find it amazing that Hogan \& Ruger's have found
two cases of D in one spectrum, when we have 2 from about 77 QSOs. 

\subsubsection{Give preference to high H~I column densities.}
There are three reasons why false identifications will be rarer when
N(H~I) is as large as possible.
First, systems with large N(H~I) are rare, so they will give 
fewer false identifications per QSO.
Second, larger $N$(H~I) means larger $N$(D~I) for a given D/H. This reduces
the chance that the D line is H, because the contaminating H would
need a large N(H~I), which is rare.
Third, if N(H~I) \aplt $10^{17}$\cm2 
then the D line will be too weak to see
in even the best Keck spectra for D/H $\simeq 2.4 \times 10^{-5}$. 
In such cases all features which look like D/H must be false identifications. 

In all three cases, the chance of false identification of an H line as D
will increase in data with lower SNR and/or lower spectral resolution,
because we have less signal to distinguish  D from H.

\subsubsection{Use Metal lines to Constrain Velocity Structure.}
Ideally metal lines  should be used to determine the velocity
structure of the absorption system. The metal lines show the
redshifts $z$ of the absorbing clouds, and both the turbulent velocity
dispersions $b_{tur}$ and temperatures $T$ of the gas in each cloud. 
These metal line parameters specify the profile of the D line, except for 
D/H and possible additional absorption by other ions, especially
H. Our two QSOs are the only ones for which the profile of D has
been fixed by metal lines. 

There must be both H and D at all redshifts seen in metal lines,
because we do not know how to remove all H and D from gas which 
contains only heavier elements and is distributed over kpc.
The fit to the H and D lines must contain gas at the redshifts
given by the metal lines, and this greatly reduces the chance of
false identification of a random H line as D.
Without metal lines, there are far more velocities at which we could
place the H and D lines.

It is always possible that the amount of H near the D line is
sufficient to require extra components in the fit.
This additional gas decreases the confidence in the identification of
D, because it broadens the D line, making it look more like an H line,
and it adds 3 degrees of freedom ($z$, $b$ \& $N$) per component, which again
increases the chance that the whole D lines is dominated by H.
For our two QSOs the D feature is completely fit by the clouds seen in the
metal lines, but for the system in  Q0014+8118 \cite{son94} additional H 
or D is required.
\section{Check List for Reliable D/H Measurements}
We summarize and conclude this discussion of the measurement of D/H in
QSO spectra with a list of some of the questions which we ask when we 
try to decide if a measurement is reliable and accurate.

\smallskip\noindent
{\bf The data:}

\smallskip\noindent
1. How many QSOs were searched to find the D/H value? There is more
chance of a false identification if only a few spectra were studied, because
systems which give secure D/H are very rare.

\smallskip\noindent
2. Was there a systematic search for D in the spectra of many QSOs?
What criteria were used to select QSOs for detailed observation?

\smallskip\noindent
3. The SNR should be high: $\simeq 100$ per 0.04\AA\ is excellent.
Lower SNR increases the chance that the D feature is contaminated with H.

\smallskip\noindent
4. The spectral resolution should be high: 8 \kms FWHM is excellent.
Lower resolution increases the chance that the D feature is contaminated with 
H.

\smallskip\noindent
{\bf The absorption system:}

\smallskip\noindent
5. How large is N(H~I)? Count only the gas which shows D.
High values are much more likely to give a real D detection.
If log N(H~I) \aplt $10^{17}$ we are unlikely to detect
D/H $\simeq 2 \times 10^{-5}$ with the best Keck data.
If D/H is low, every claim of D in a system with low N(H~I)
must be contaminated.

\smallskip\noindent
6. How well is N(H~I) determined? If from a Lyman limit alone, is it know 
that there are no other systems which could account for the Lyman
continuum absorption? If from Lyman line, the more lines the better.
One saturated line is inadequate in low SNR data, but may be acceptable
with high SNR and high resolution.
Ideally N(H~I) is from many Lyman lines and the Lyman continuum.

\smallskip\noindent
7. How certain is it that there is a feature at the position of D?

\smallskip\noindent
8. How was the position of D determined?
If from the D line itself, then there is a large chance of H contamination.
Ideally from metal lines and narrow high order Lyman lines.

\smallskip\noindent
9. How many components were required to fit D?
The fewer the better.

\smallskip\noindent
10. Are additional H components needed to fit around D? This is bad
and increases the chance of contamination.

\clearpage
\begin{table*}
\begin{center}
\caption{Reliability of D/H Measurements in QSO Spectra.}
\begin{tabular}{|lccccccc|}
\hline   
  & & & & & & & \\
 & & \multispan{4}{Deuterium~System} & & \\
 Criterion & (A) & (B) & (C) & (D) & (E) & (F) & (G) \\
 & & & & & & & \\
\hline   
 & & & & & & & \\
 Many QSOs searched? & 7  & 7  & 3  & 3  & 3  & 3  & 0  \\
 Systematic D/H search? & 10 & 10 & 2  & 0  & 0  & 0  & 0  \\
 High SNR? & 5  & 8  & 4  & 4  & 3  & 2  & 9  \\
 High spectral resolution? & 10 & 10 & 10 & 10 & 5  & 5  & 10 \\
 High N(H~I)? & 6  & 8  & 2  & 8  & 2  & 10 & 0  \\
 Small $\sigma $ for N(H~I)? & 6  & 7  & 8  & 2  & 7  & 0  & 4  \\
 Definite feature at D? & 10 & 10 & 7  & 0  & 7  & 0  & 10 \\
 D $v$ from other ions? & 10 & 10 & 10 & 8  & 7  & 10 & 0  \\
 Few components to fit D? & 5  & 5  & 3  & 0  & 10 & 0  & 5  \\
 Extra H to fit D? & 0  & 0  & -5 & 0  & 0  & -5 & 0  \\
 Component $v$ from where? & 10 & 10 & 5  & 10 & 0  & 0  & 0  \\
 How good is fit to data? & 10 & 10 & 5  & 2  & 5  & 8  & 10 \\
Is fit unique? &  0 & 0  & 0  & -5 & 0  & -5 & 0  \\
 Is $b$(D) known? &  5 & 5  & 7  & 5  & 0  & 5  & 5  \\
Was $b$(D) predicted?  & 7 & 7  & 0  & 7  & 0  & 5  & 0  \\
Agreement on $b$(D)? & 7 & 7  & 0  & 7  & 0  & 3  & 0  \\
Metal abundance low? & 10 & 10 & 10 & 7  & 0  & 5  & 0  \\
Additional evidence? & 5 & 5  & 0  & 0  & 0  & 5  & 0  \\
H contamination corrected? & 10 & 10 & 0  & 0  & 0  & 0  & 0  \\
\hline
Total & 133 & 139 & 71 & 58 & 49 & 51 & 53 \\
(out of 170) &  & & & & & & \\
\hline
\end{tabular}
\end{center}
\end{table*}

\smallskip\noindent
11. How were the velocities of the components determined?
If from D, then there is a high chance of contamination.
Ideally from metal lines and narrow high order Lyman lines.

\smallskip\noindent
12. How good is the fit? It must be excellent in the critical regions which
distinguish different values of D/H.

\smallskip\noindent
13. Is the fit unique? Do different fits give different D/H?

\smallskip\noindent
14. Was the $b$ value of D measured? If not, why not? Lack of a measurement is
a sign of a non-unique fit, poor data, or strong blending, all of which
favor contamination by H.

\smallskip\noindent
15. Was the $b$ value of D predicted?
Whenever metal lines are seen the temperature and $b_{tur}$ should be 
measured and used to give a prediction.
If $b$ is not measured, then $b$ should lie between $b$(H) and
$b$(H)/$\sqrt{2}$.
Gas which is suitable for the detection of D will be cool, and will have
low $b_{tur}$, but
thermal motions will probably dominate the $b$ values of D and H.

\smallskip\noindent
16. How tight is the agreement between the predicted and measured $b$(D)?

\smallskip\noindent
17. Was the metal abundance measured? If not, why not? It could be high.

\smallskip\noindent
18. Is there additional evidence that the fit to the system is reasonable,
such as standard element abundance ratios and reasonable temperatures
(few $10^{4}$ for ionized gas)?

\smallskip\noindent
{\bf The analysis:}

\smallskip\noindent
19. Was there a Monte Carlo correction for \Lya forest lines at the
position of D. If not, why not? Such corrections are hard to make
when fits are poor or ambiguous.

In Table 3 we include subjective estimates of how well each absorption
system (see Table 2) performs on the 19 questions.
We assign a score of 10 for an excellent indication of good D/H, zero
for no information, and negative numbers for problems which make
D/H uncertain.
Questions 19 is the key, since it summarizes the probability
that the D line is contaminated.
The two absorption systems discussed in this paper are the only ones to pass
all of these tests. The others fail many tests, and
for this reason they are likely to be strongly contaminated by hydrogen.
Note that the absorption in QSO 1422+2309 (G), which we think is completely
contaminated with H, ranks as well or better than many others.
\section{Acknowledgments}
We thank David Kirkman for the spectrum showing the Lyman alpha forest,
Michael Turner for asking what happens if the two clouds have different D/H,
Richard Mushotzky for asking why the relative velocities of the
two clouds are low, Patrick Diamond for discussions of turbulence, and
Sandra Faber for asking if we can model H~I gas associated with the D/H
absorber.  We are especially grateful to George Fuller for suggesting that
we write a detailed description of how we measure D/H, and to
Taka Kajino for organizing and hosting the very special meeting in Atami.

\end{document}